\documentclass[twocolumn,showpacs,pra,aps,graphicx,preprintnumbers,amsmath,amssymb]{revtex4}

\usepackage{dcolumn}
\usepackage{bm}

\usepackage{epic}
\usepackage{eepic}
\usepackage{graphicx}

\renewcommand{\vec}[1]{{\bf{#1}}}

\begin{document}

\title{Reconstruction of the phase of matter-wave fields using a momentum
resolved cross-correlation technique}
\author{D. Meiser}
\author{P. Meystre}
\affiliation{Optical Sciences Center, The University of Arizona,
Tucson, Arizona 85721}


\begin{abstract}
We investigate the potential of the so-called XFROG
cross-correlation technique originally developed for ultrashort
laser pulses for the recovery of the amplitude and phase of the
condensate wave function of a Bose-Einstein condensate. Key
features of the XFROG method are its high resolution,
versatility and stability against noise and some sources of
systematic errors. After showing how an analogue of XFROG can be
realized for Bose-Einstein condensates, we illustrate its
effectiveness in determining the amplitude and phase of the wave
function of a vortex state. The impact of a reduction of the
number of measurements and of typical sources of noise on the
field reconstruction are also analyzed.
\end{abstract}

\maketitle

\label{introduction}

Many of the remarkable properties of atomic Bose-Einstein
condensates (BECs) originate from the fact that in those systems a
single wave-function is occupied by a macroscopic number of
particles
\cite{Ketterle:BEC1995,Hulet:BEC1995,Cornell:BEC1995,Ketterle:BEC_coherence,Ketterle:BEC_interference}.
The amplitude of this complex wave function corresponds to the
atomic density and can comparatively easily be measured using for
example absorption or phase contrast imaging
\cite{You:QFT_III,You:absorption_imaging,Politzer:Born,Politzer:Bose_stimulated_scattering,Ketterle:BEC_review}. The phase of the wave function, on the
other hand, is relatively hard to measure directly. Still, its
measurement is essential for the characterization of many
quantum-degenerate atomic systems, such as e.g. rotating BECs with
vortices \cite{Matthews:Vortices}.

A situation in many ways similar is encountered in the context of
ultrashort laser pulses of a few optical cycles duration. The
characterization of these pulses, which are of considerable
interest in both fundamental science and applications, requires
likewise the knowledge of both amplitude and phase.

The measurement of the time-dependent phase of ultrashort laser
pulses has found a solution that is in many respects optimal in
the so-called Frequency-Resolved Optical Gating (FROG) methods
\cite{Chilla_Martinez,Kane:FROG1,DeLong:SHG_FROG,Trebino:JOSA}.
These methods, which can be adapted to many different situations,
offer very high resolution and precision, are stable against
noise, and can even compensate for or detect some sources of
systematic errors. Single-shot measurements are straightforward
\cite{Kane:Single_shot}, and measurements of fields with less than
one photon per pulse on average have also been successfully
demonstrated. Almost every aspect of the FROG-methods has been
studied in depth. Good starting points for accessing the wealth of
research literature on that subject are the review article
\cite{Trebino:FROG_review} and Ref. \cite{Trebino:FROG_book}.

The similarity of the phase-retrieval problems for ultrashort
pulses and atomic condensates naturally leads one to ask whether the
optical FROG techniques can be extended to the atom-optical
domain. This paper answers this question in the affirmative by
demonstrating how a specific version of the general FROG method,
called cross-correlation FROG (XFROG), can easily be translated to
the atom-optical case
\cite{Linden:Downconversion,Linden:XFROG,Linden:XFROG_book}.

Section \ref{frog_basics} briefly reviews the fundamentals of
XFROG phase retrieval and discusses how it can be adapted to
cold-atom scenarios. As an illustration, section
\ref{vortex_reconstruction} shows the reconstruction of the
spatial phase of a rotating BEC with a vortex. We also investigate
the sensitivity of the reconstruction to several sources of noise
characteristic of cold atoms, as well as the impact of a reduction
in the number of measurements on the field retrieval.

\section{\label{frog_basics}
Principles of the XFROG method}

\begin{figure}
\includegraphics{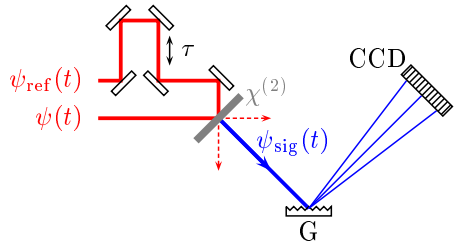}
\caption{(Color online) Basic experimental setup for the XFROG method. G is a
diffraction grating and the spectrum of the signal field is recorded with the
CCD array.} \label{xfrog_setup}
\end{figure}

The basic experimental setup for the field reconstruction of an
ultrashort laser pulse by means of the XFROG method is shown in
Fig. \ref{xfrog_setup}. For simplicity we assume that only one
direction of polarization of the electrical field needs to be
considered. The unknown field $\psi(t)$ is mixed with a known
reference field $\psi_{\rm ref}(t)$ in a nonlinear crystal with
$\chi^{(2)}$ nonlinearity after a variable delay $\tau$. The
fields $\psi_{\rm ref}$ and $\psi$ should be of roughly comparable
duration which means in practice that the pulse durations can
differ by up to about an order of magnitude. The sum frequency
signal is
\begin{equation}
\psi_{\rm sig}(t,\tau)\propto \psi(t)\psi_{\rm ref}(t-\tau).
\label{optfrogsignal}
\end{equation}
The form of that field shows that the reference pulse acts as a
gate for the pulse, hence the acronym XFROG. The signal $\psi_{\rm
sig}$ is then spectrally analyzed. The resulting spectrum,
\begin{eqnarray}
I_{\rm XFROG}(\omega,\tau)&\equiv& \left|
\int d t e^{-i\omega t}\psi_{\rm sig}(t,\tau)\right|^2\nonumber \\
&=&\left|\int dt e^{-i\omega t}\psi(t)\psi_{\rm
ref}(t-\tau)\right|^2, \label{xfrog-def}
\end{eqnarray}
is the key quantity. It contains enough information to reconstruct
amplitude and phase of the pulse $\psi(t)$.

To see how this works \footnote{In the following we denote
functions and their Fourier transforms by the same symbol to keep
the notation simple. Which one is meant will be clear from the function
arguments.}, we first note that it is sufficient to find $\psi_{\rm
sig}(t,\tau)$ since $\psi(t)$ can then be obtained by simply
integrating over $\tau$ --- up to a multiplicative constant that
can be determined from the normalization of $\psi$. Thus the
problem is cast in the form of a two-dimensional phase retrieval
problem. Problems of this type have been studied for decades in
the context of image restoration, see e.g.
\cite{Fienup:Algorithm_comparison,Yudilevich:generalized_proj}.

The field $\psi_{\rm sig}$ can be found from $I_{\rm XFROG}$ by
means of the method of generalized projections
\cite{DeLong:Improved_algorithm,Yudilevich:generalized_proj,DeLong:Pulse_retrieval_GP}.
From Eqs. (\ref{optfrogsignal}) and (\ref{xfrog-def}) it is clear
that $\psi_{\rm sig}$ must simultaneously belong to the two sets
\footnote{For ease of notation we do not specify the function
spaces our functions are elements of. All functions of importance
to us are such that the expressions we write down are meaningful.}
\begin{equation}
O=\left\{f(t,\tau) | f(t,\tau)=g(t) \psi_{\rm ref}(t-\tau)\text{ for some } g\right\}
\label{Odef}
\end{equation}
and
\begin{equation}
F=\left\{f(t,\tau) | \left|\int dt e^{-i\omega
t}f(t,\tau)\right|^2 =I_{\rm XFROG}(\omega,\tau)\right\}.
\label{Fdef}
\end{equation}
Hence, the solution is a field that satisfies the two constraints
\begin{equation}
\psi_{\rm sig}(t,\tau)\in O\cap F.
\end{equation}

\begin{figure}
\includegraphics{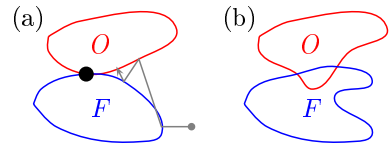}
\caption{(Color online) Illustration of the constraint sets $O$
and $F$ (a) in the case where $O$ and $F$ are convex and their
intersection contains exactly one element and (b) in the
non-convex case with more than one element in the intersection.
Fig. (a) also shows a typical trajectory of the generalized
projections algorithm.} \label{intersection}
\end{figure}

Figure \ref{intersection}a suggests that one can find the solution
to this problem, called a feasibility problem in mathematics and
especially in optimization theory, by iteratively projecting onto
the two constraint sets $O$ and $F$. For closed convex sets with
exactly one point of intersection, this method always leads to a
unique solution. In our case, though, the constraint sets are not
convex and their intersection consists of more than one function,
as illustrated in Fig. \ref{intersection}b. Hence, the algorithm
of generalized projections is not guaranteed to converge for every
initial guess and even if it does the solution is not unique.
Still, in practice it converges for a vast majority of initial
guesses and the ambiguity in the solution is physically
reasonable. (In particular, XFROG determines the field $\psi_{\rm
sig}$ up to a constant phase.) On those rare occasions when the
algorithm does not converge for a particular initial guess, this
non-convergence is revealed by a large distance of the fixed point
from the constraint sets in a sense that will be made precise
below. Then one can simply restart the reconstruction algorithm
with a different initial guess.

Specifically, $\psi_{\rm sig}$ is found from $I_{\rm XFROG}$ as follows:

\begin{enumerate}

\item Initialize $\tilde{\psi}_{\rm sig}^{(0)}(\omega,\tau)$ with
random numbers for its real and imaginary part. \item The
projection onto the constraint set $F$ is accomplished by setting
\begin{equation}
\psi_{\rm sig}^{(n)}(\omega,\tau)= \frac{\tilde{\psi}_{\rm
sig}^{(n-1)}(\omega,\tau)}{|\tilde{\psi}_{\rm
sig}^{(n-1)}(\omega,\tau)|} I_{\rm FROG}^{1/2}(\omega,\tau),
\end{equation}
which clearly guarantees that $\psi_{\rm sig}^{(n)}$ 
satisfies condition (\ref{Fdef}).

\item Inverse Fourier transform $\psi_{\rm
sig}^{(n)}(\omega,\tau)$ with respect to the first argument to
find
\begin{equation}
\psi_{\rm sig}^{(n)}(t,\tau)=\int\frac{d\omega}{2\pi}e^{i\omega t}
\psi_{\rm sig}^{(n)}(\omega,\tau).
\end{equation}

\item
Determine $\psi^{(n)}(t)$ such that
\begin{equation}
Z=\int dtd\tau \left|\psi_{\rm sig}^{(n)}(t,\tau)
-\psi^{(n)}(t)\psi_{\rm ref}(t-\tau)\right|^2
\label{frogerror}
\end{equation}
becomes a minimum. With $\psi^{(n)}(t)$ determined this way form
\begin{equation}
\tilde{\psi}_{\rm sig}^{(n)}(t,\tau) =\psi^{(n)}(t)\psi_{\rm
ref}(t-\tau),
\end{equation}
which is closer to the set $O$ than $\psi_{\rm sig}^{(n)}(t,\tau)$.
\item Use the Fourier transform
\begin{equation}
\tilde{\psi}_{\rm sig}^{(n)}(\omega,\tau)=
\int dt e^{-i\omega t} \tilde{\psi}_{\rm sig}^{(n)} (t,\tau)
\end{equation}
as a new input in step 2, and iterate until the error $Z$ in Eq.
\eqref{frogerror} becomes sufficiently small.
\end{enumerate}
Upon exit from the algorithm, $\psi^{(n)}$ is the retrieved field.
It turns out that in step 4 a simple line minimization along the
gradient of $Z$ with respect to $\psi^{(n)}$ is sufficient. A good
starting point for that minimization is the minimizer
$\psi^{(n-1)}$ of the previous step.

An important characteristic of the XFROG method is its robustness,
which results from two main reasons: First, the XFROG signal
contains a high degree of redundancy: If $I_{\rm XFROG}$ is
measured on a grid of size $N\times N$ the $2N$ unknowns of the
field $\psi(t)$ -- its real and imaginary parts -- are retrieved
from $N^2$ measured values. The XFROG algorithm makes use of this
high degree of redundancy to yield a highly stable pulse
retrieval. Second, the functional form of $\psi_{\rm sig}$ is very
restrictive in the sense that a randomly generated $\psi_{\rm
sig}$ will not correspond to any physical pulse $\psi$. Thus, if
the XFROG signal has been deteriorated by systematic errors the
XFROG algorithm will not converge for any initial guess and one
can conclude that the data is corrupted. Another advantage of
XFROG is the extremely high temporal resolution that is achieved
by making use of the Fourier domain information. Instead of being
determined by the length of the reference pulse, it is essentially
given by the response time of the non-linear medium.

We now turn to the central point of this paper, which is to adapt
the XFROG method to the characterization of matter-wave fields. We
consider specifically the case of atomic bosons, with two internal
states denoted by $\uparrow$ and $\downarrow$. The atoms are assumed to
be initially in a pure Bose-Einstein condensate (BEC) at
temperature $T=0$, with all atoms in internal state $\downarrow$.
The corresponding atomic field operator is
\begin{equation}
\hat{\psi}(\vec{r})=\hat{c}_\downarrow\psi(\vec{r}),
\end{equation}
where $\psi(\vec{r})$ is the condensate wave function and
$\hat{c}_\downarrow$ is the bosonic annihilation operator for an
atom in the condensate.

The states $\uparrow$ and $\downarrow$ are coupled by a spatially
dependent interaction of the generic form
\begin{equation}
V_J =\int d^3r {V}(\vec{r}-\vec{R})
\hat{\psi}_\uparrow^\dagger(\vec{r})
\hat{\psi}_\downarrow(\vec{r}) +H.c. \label{outcoupling}
\end{equation}
that is switched on at time $t=0$. This could be provided for
example by a two-photon Raman transition, with $V$ then being
proportional to the product of the mode functions of the two
lasers driving the transition.

For short enough times the $\uparrow$-component of the atomic
field is then
\begin{equation}
\hat{\psi}_\uparrow(\vec{r})\propto
\tilde{V}(\vec{r}-\vec{R})\hat{\psi}_\downarrow(\vec{r}).
\end{equation}
The resulting momentum distribution of the $\uparrow$-atoms is
\begin{eqnarray}
n(\vec{q},\vec{R})&\propto&\int d^3rd^3r^\prime
e^{i\vec{q}\vec{r}}e^{-i\vec{q}\vec{r}^\prime}
\left\langle \hat{\psi}_\uparrow^\dagger (\vec{r})\hat{\psi}_\uparrow(\vec{r}^\prime)
\right\rangle\\
&\propto&\int d^3rd^3r^\prime
e^{i\vec{q}\vec{r}}e^{-i\vec{q}\vec{r}^\prime}\nonumber\\
&&\times
\psi^*(\vec{r})V^*(\vec{r}-\vec{R})
\psi(\vec{r}^\prime)V(\vec{r}^\prime-\vec{R})\\
&=&\left|\int d^3 r e^{-i\vec{q}\vec{r}}
\psi(\vec{r})V(\vec{r}-\vec{R})\right|^2.
\label{atomfrogsignal}
\end{eqnarray}
When measured as a function of the shift $\vec{R}$ between
coupling and condensate, the momentum distribution of the
$\uparrow$-atoms is therefore an XFROG signal, with the role of
the reference field being played by the space dependent coupling
strength. Hence, the XFROG algorithm can be used to fully recover
the field $\psi$, $n({\bf q}, {\bf R})$ being measured using for
example absorption imaging after free expansion.

In some sense, the situation for atoms is simpler than for
ultrashort laser pulses. This is because a gate function of size
comparable to the condensate size, such as e.g. the coupling
strength $V(\vec{r})$, is readily available. While in the case of
photons the only possible gate is another ultrashort laser pulse,
and hence one must use nonlinear mixing to obtain a signal
suitable for XFROG, for atoms one can use processes that are
linear in the atomic field. Furthermore, since the interactions
between cold atoms tend to be stronger and offer a richer variety
than is the case for light, a wide choice of processes can be
employed to generate signals that can be used for FROG algorithms.
Specific examples include $s$-wave interactions, the atom-optics
analog of optical cubic nonlinearities, as well as the recently
demonstrated coherent coupling of atoms to molecules, which
corresponds to optical quadratic nonlinearities
\cite{Wynar:MoleculePa,Inouye:MoleculeFR,Donley:MoleculeBEC,Duerr:MoleculeRP,Greiner:MoleculeBEC,Jochim:Li2}.

A fully three-dimensional XFROG scheme as suggested by Eq.
(\ref{atomfrogsignal}) is hard to realize in practice. First, it
requires a very large number of measurements. Ten different shifts
in each direction correspond to a total of 1000 runs of the
experiment. The large amount of data necessary for the fully
three-dimensional scheme also poses serious challenges to the
numerical reconstruction algorithm as far as computer memory and
time are concerned. Second, in time-of-flight absorption imaging
one typically measures the column-integrated density so that the
three-dimensional momentum distribution is not directly
accessible. It appears therefore preferable to limit the
reconstruction to have a two-dimensional scheme.

It can be easily verified that if the field $\psi$ and the
coupling strength can be factorized as
\begin{equation}
\psi(\vec{r})=f(z)\psi(x,y),\quad V(\vec{r})=h(z)V(x,y),
\label{factorization}
\end{equation}
where $z$ is the direction along which the imaging is done, Eq.
\eqref{atomfrogsignal} remains valid if we interpret $\vec{r}$ and
$\vec{R}$ as two-dimensional vectors in the plane perpendicular to
$z$ and $\vec{q}$ as a corresponding two-dimensional momentum.
Many fields of practical interest can at least approximately be
written in the form Eq. \eqref{factorization}. For example, the
two-photon Raman coupling mentioned earlier is of that type
provided that the lasers are directed along $z$ and the Rayleigh
length is much longer than the extend of the atomic cloud in that
direction. As a concrete example we demonstrate in the next
section how the matter wave field of a rotating BEC with a single
vortex \cite{Matthews:Vortices} can be directly reconstructed using the
XFROG method.

\section{\label{vortex_reconstruction}Reconstruction of a vortex-field}

\begin{figure*}
\includegraphics[width=0.95\columnwidth]{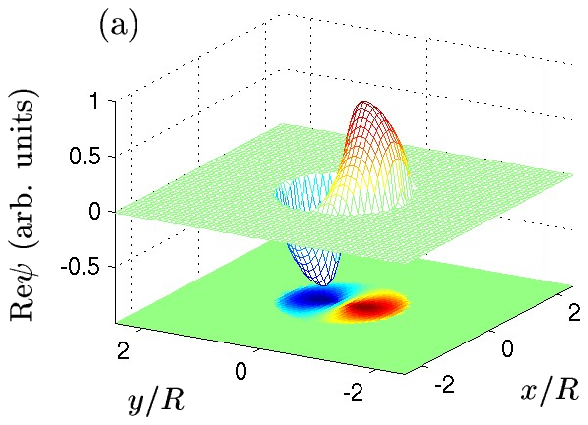}
\includegraphics[width=0.95\columnwidth]{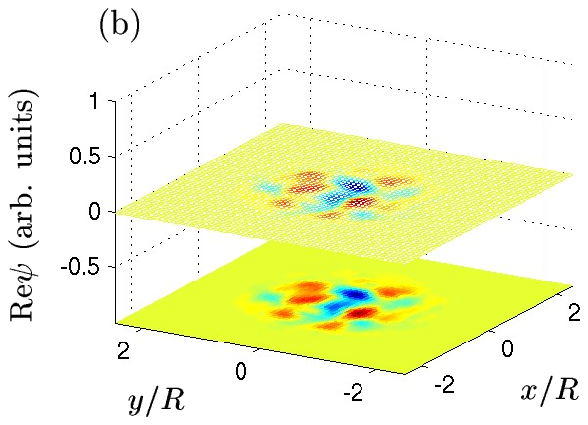}
\includegraphics[width=0.95\columnwidth]{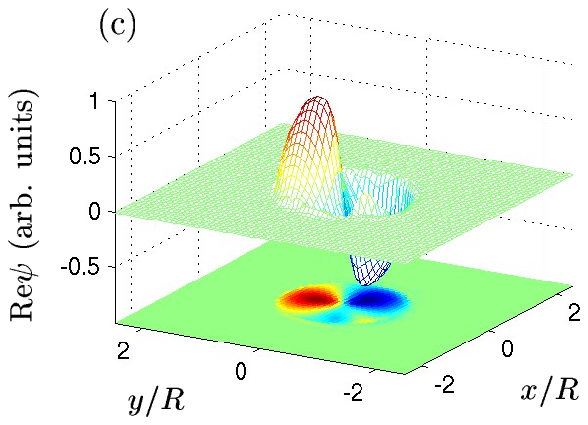}
\includegraphics[width=0.95\columnwidth]{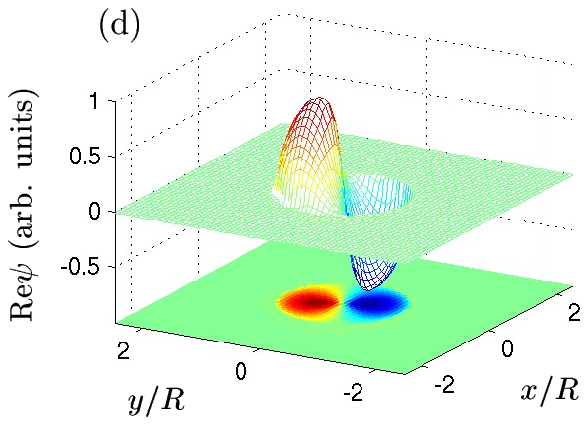}
\caption{(Color online) Reconstruction of a vortex state on a grid of
$64\times64$ points.  Figure (a) shows the real part of the original state and
figures (b)-(d) show the initial guess and the reconstructed state after 50 
and 100 iterations, respectively.}
\label{vortexreconstructionfig}
\end{figure*}

We consider a BEC at temperature $T=0$ in a spherical trap. We
assume that the Thomas-Fermi approximation holds and that the
healing length $\xi = (8\pi a n_0)^{-1/2}$, with $a$ the $s$-wave
scattering length and $n_0$ the atomic density at the center of
the trap, is much smaller than the Thomas-Fermi Radius $R=(15 N_a
a/a_{\rm osc})^{1/5}a_{\rm osc}$ of the cloud. Here, $N_a$ is the
number of atoms in the condensate and $a_{\rm osc}$ is the
oscillator length of the atoms in the trap. Under these
assumptions the structure of the vortex core is essentially the
same as that of a vortex in a uniform BEC and its wave function
can to a good approximation be written as \cite{Smith_Pethick}
\begin{equation}
\psi(r,\varphi,z)=f(z)\frac{r/R}{\sqrt{2(\xi/R)^2+(r/R)^2}}
\sqrt{1-(r/R)^2}e^{i\varphi},
\label{vortexpsi}
\end{equation}
where we have used cylindrical coordinates with the vortex core at
the symmetry axis.  As discussed above, the $z$-dependence of the
wave-function is unimportant and we will not regard it any
further. In the following we use $\xi/R=0.1$, but none of our
results depend strongly on this ratio as long as $\xi/R\ll 1$. The
real part of the wave-function \eqref{vortexpsi} is shown in Fig.
\ref{vortexreconstructionfig}a.

As an interaction Hamiltonian, or `reference field' in the
language of XFROG, we use Eq. \eqref{outcoupling} with the Gaussian
\begin{equation}
V(r)=e^{-(r/w)^2},
\label{referencepsi}
\end{equation}
the $z$-dependence being again irrelevant for our purposes.

\begin{figure}
\includegraphics[width=0.95\columnwidth]{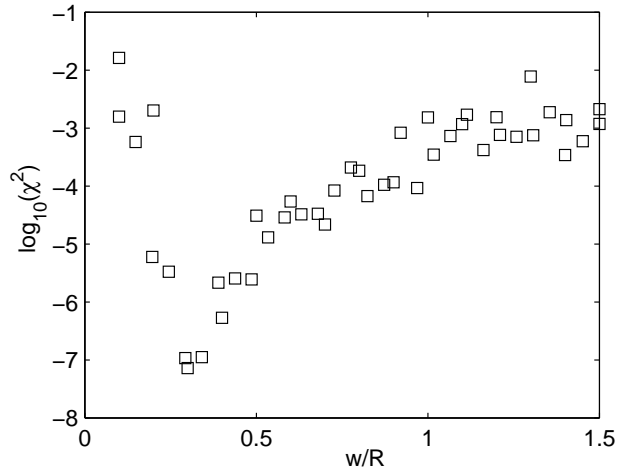}
\caption{$\chi^2$ error of the reconstructed vortex field after
100 iterations as a function of the width $w$ of $V$.}
\label{widthseries}
\end{figure}

We have simulated an XFROG signal on a grid of $64\times 64$ points
using $\psi$ of Eq. \eqref{vortexpsi} and the reference field
\eqref{referencepsi}, and applied the XFROG algorithm to
reconstruct the field. In order the determine the optimal width of
$V$ we have repeated this procedure for various values of $w$. For
each run we have determined the $\chi^2$-error per degree of
freedom \footnote{For $\chi^{2}$ of Eq. \eqref{chitwo} to be
useful as a measure of the error it is necessary to determine the
arbitrary total phase of $\psi^{(n)}$ such that $\chi^{2}$ becomes
a minimum.}
\begin{equation}
\chi^2 = \frac{1}{2 N^2}\sum_{i,j}|\psi^{(n)}(\vec{r}_{ij})
-\psi(\vec{r}_{ij})|^2, \quad i,j\text{ grid points},
\label{chitwo}
\end{equation}
after 100 iterations. Here, $N$ is the number of grid points in
one direction. The results of these simulations are summarized in
Fig. \ref{widthseries}. They show that the XFROG algorithm works
rather well for a wide range of widths provided that they are
comparable to the size of the condensate.
As a rule of
thumb, for our data $\chi^2$ errors smaller than $10^{-2}$ mean
that the algorithm has qualitatively recovered the original
field.
The comparatively poor quality of the retrieved fields for larger widths is
mainly due to unphysical correlations across the boundaries arising from the
periodic boundary conditions that we are using. These effects can in principle
be avoided by using a larger grid. The best results were obtained for widths of
$w\approx 0.3R$ and in all that follows we use $w=0.35R$. Larger widths tend to
render the algorithm more stable, in the sense that it will converge to the
correct solution for more initial guesses, while narrower reference fields
result in faster convergence, but for fewer initial guesses.

\begin{figure}
\includegraphics[width=0.9\columnwidth]{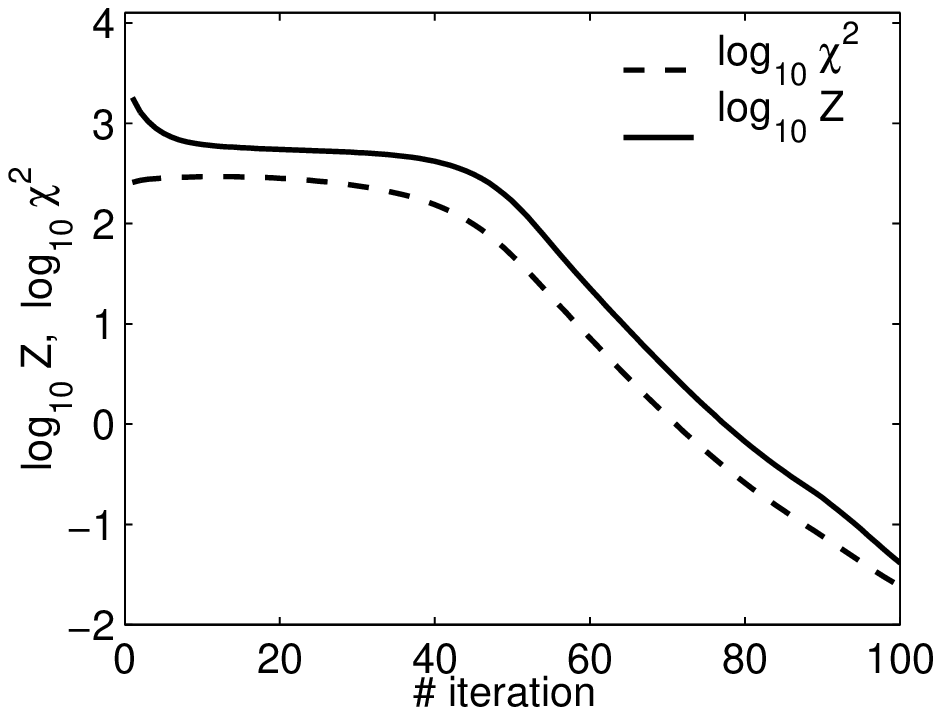}
\caption{FROG error $Z$ and field error $\chi^2$ for the simulation run
of Fig. \ref{vortexreconstructionfig}.}
\label{FROGerrorfig}
\end{figure}

Several stages of the reconstruction algorithm are shown in
figures \ref{vortexreconstructionfig}(b)-(d). (The field
reconstruction in this example takes about 30 minutes on a Pentium
4 CPU and uses approximately 400 MB of memory.) We show only the
real part of the field, as the imaginary part shows a similar
degree of agreement. The XFROG error $Z$ of Eq. \eqref{frogerror}
for the same simulation run is shown in Fig. \ref{FROGerrorfig} as a function
of the number of iterations. Also shown is the deviation from the reconstructed
field from the original field $\chi^2$. The figure shows that the algorithm
converges exponentially after some initial stagnation.  It also shows that,
after the ambiguity in the total phase has been taken into account, the XFROG
error $Z$ is a good measure of the actual discrepancy between reconstructed and
original field.  This is important for real-life applications since in practice
one does not know the original field so that $\chi^2$ cannot be calculated and
only the XFROG error is accessible.

Time of flight absorption imaging is a comparatively easy way to
obtain many data points in the Fourier domain, but it is
cumbersome to obtain the data sets for different shifts $\vec{R}$
because each such set requires a new run of the experiment. Thus a
natural question is the sensitivity of the algorithm to the number
of displacements for which a measurement is performed.

\begin{figure}
\includegraphics[width=0.95\columnwidth]{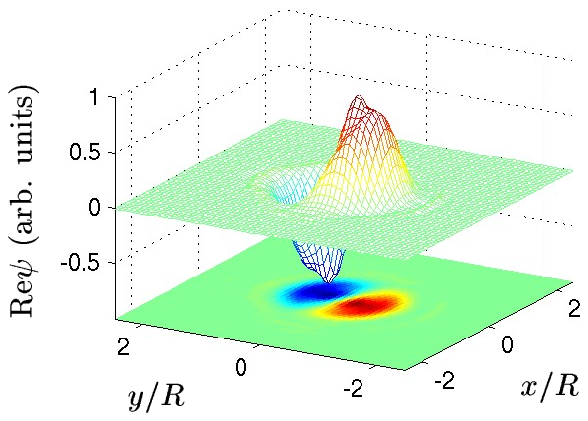}
\caption{(Color online) Reconstruction of the field of Fig. \ref{vortexreconstructionfig} with the XFROG signal simulated on a grid of $10\times 10$ and
interpolated on a grid of $64\times 64$ after 100 iterations.}
\label{interp_fig}
\end{figure}

To answer this question we have calculated XFROG signals $I_{\rm
XFROG}$ on grids of dimension $n\times n$, interpolated them onto
a larger grid of dimension $N\times N$ using cubic splines, and
applied the XFROG algorithm to the resulting XFROG signals. Fig.
\ref{interp_fig} shows an example for $n=10$. The
recovered field shows good agreement with the original field, with
differences in some details, e.g. near the maxima, resulting from
the smoothing property of the interpolation with splines.

\begin{figure}
\includegraphics[width=0.9\columnwidth]{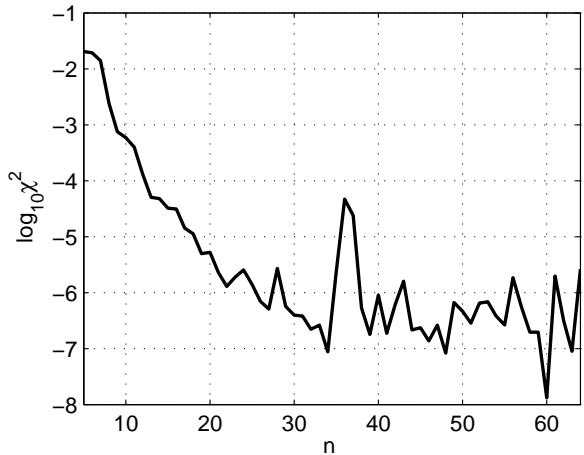}
\caption{$\chi^{2}$ error of the recovered field after 100
iterations as a function of the number of measured data points $n$
(same parameters as in figures \ref{vortexreconstructionfig} and
\ref{FROGerrorfig}).} \label{interpseries}
\end{figure}

To quantitatively characterize the dependence of the success of
the field recovery on the number of measurements we have evaluated
$\chi^2$ of the recovered field after 100 iterations as a function
of $n$. The result is shown in Fig. \ref{interpseries}. While the
discrepancy between the recovered field and the original field
grows as expected as the number of measurements decreases, the
field is qualitatively retrieved down to $n=7$. 
For even smaller $n$ the field reconstruction fails. In
those cases the algorithm is honest enough to admit its failure by
generating a large residual XFROG error. Nonetheless, the
conclusion is that a rather limited number of measurements is
sufficient to accurately measure the field.

We remark that these considerations are actually overly
pessimistic. In reality one often knows the size of the atomic
cloud and the XFROG signal is clearly zero if there is no overlap
between coupling $V$ and atomic field $\psi$. Thus, many data
points in the XFROG signal can be padded with zeros. Even those
zeros contain usable information for the XFROG algorithm because
they encode a support constraint for the recovered field. Support
constraints, together with Fourier domain information, are widely
used in image reconstruction for astronomic imagery and under
certain circumstances they characterize an image completely.
Furthermore, among the remaining non-zero measurements some will
give excessively small signals. In practice one can replace these
measurements by other displacements that give a stronger signal.

In an actual experiment the measurement of an XFROG signal will be
subject to several sources of noise. In that case the intersection
of the sets $O$ and $F$ in Fig. \ref{intersection} will in general
be empty, and the XFROG algorithm determines a field that has the
correct XFROG signal, i.e. belongs to the set $F$, and at the same
time minimizes the distance to the set $O$ as measured by $Z$.

The XFROG signal Eq. (\ref{frogerror}) is insensitive to
fluctuations in the total phase of the atomic field $\psi$ from
one run of the experiment to the other, i.e.  for different
$\vec{R}$ --- a consequence of the non-interferometric character
of FROG methods in general. Hence it is sufficient to study the impact of
uncertainties in the measurement of the displacements $\vec{R}$
themselves and in the total intensity of the XFROG signal from shot
to shot. The latter can arise from variations of the number of
atoms in the original condensate, from fluctuations in the
coupling strength, and from fluctuations in the interaction times.
We have simulated XFROG signals with Gaussian fluctuations in
$\vec{R}$ and in shot-to-shot total intensity $I$ with variances
$\Delta\vec{R}$ and $\Delta I$, respectively. Note that with our
choice of normalization for $\psi$ and $V$ the total intensity of
the XFROG signals is of order one so that we can treat absolute
errors and relative errors in $I$ as being the same.

\begin{figure}
\includegraphics[width=0.9\columnwidth]{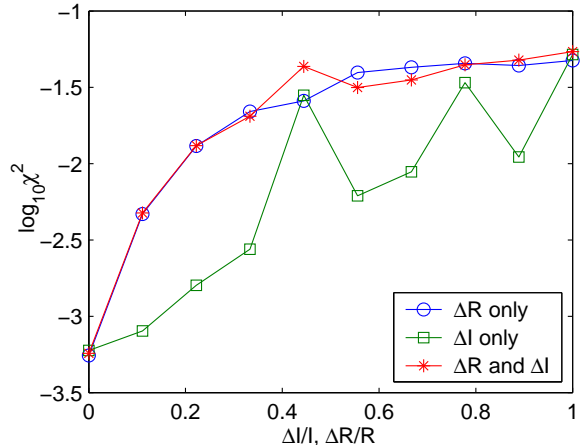}
\caption{(Color online) $\chi^{2}$ error of the recovered field
after 100 iterations as a function of several noise sources.}
\label{errorfig}
\end{figure}

We have applied the XFROG algorithm to these contaminated signals
and we have measured $\chi^2$ after 100 iterations.  To be more
realistic, we have simulated the XFROG signals only on a grid of
$10\times 10$ and interpolated onto a grid of $64\times 64$ as
above.  We have simulated the situation with errors in $\vec{R}$
only, errors in $I$ only, and errors in both $\vec{R}$ and $I$.
The results are summarized in Fig. \ref{errorfig}. We were able
to qualitatively reconstruct the atomic field up
to relative errors as large as $\Delta R/R\approx 0.3$ and $\Delta I/I =1$. The
algorithm was found to be significantly less sensitive to
uncertainties in the total intensity than to uncertainties in the
displacements. Even XFROG signals with $\Delta I/I=1$ yield
recovered fields of very high quality. The relatively high residual
errors for $\Delta I/I \approx 0.4$ and $\Delta I/I \approx 0.8$ come about
because the XFROG algorithm did not find the global minimum of $Z$ for
the particular initial guess. Restarting the algorithm with a different
initial guess (but with the exact same contaminated XFROG signal) lead to
residual errors that nicely interpolate the the other data points.

\begin{figure}
\includegraphics[width=0.95\columnwidth]{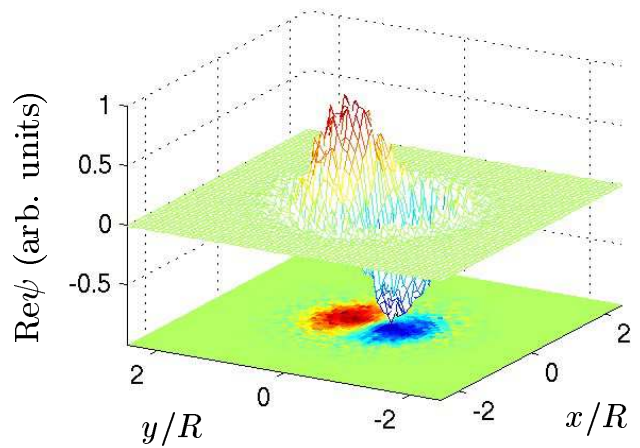}
\caption{(Color online) Same as Fig. \ref{interp_fig} but with
simulated noise $\Delta R/R=0.1$ and $\Delta I/I=0.1$ in the XFROG
signal.}
\label{vortex_noise}
\end{figure}

Figure \ref{vortex_noise} shows the retrieved field
after 100 iterations for $\Delta R/R=0.11$ and $\Delta I/I=0.11$.
The broad features of the original field are clearly reproduced
and the noisy fine structure is exclusively due to the
fluctuations in $\vec{R}$ and could be smoothed out by convolution
with a Gaussian of width comparable to the healing length. Thus,
although the error in the retrieved field increases with
increasing noise, we conclude that the algorithm is not very
sensitive to noise and yields at least reliable qualitative
information. In a sense, the field as recovered by the XFROG
method contains much less noise than the input data. This is
reminiscent of the situation encountered in image reconstruction
e.g. in astronomical applications. And it was exactly for the
purpose of removing noise from images by using a mixture of
position space and Fourier space information that many of the
methods described in this paper were first discussed.

In summary we have shown how the powerful XFROG method of
ultrafast optics can be adapted to phase measurement applications
in ultra-cold atomic systems. In particular, this technique 
is capable of correctly recovering a vortex state and it is robust
against noise, with a number of actual measurements that can
be reduced to experimentally feasible values.

An interesting question for the future is how the XFROG method can be used to
study situations were there is no well defined phase of the atomic field for
fundamental reasons. For example the XFROG technique could be a valuable tool
for diagnosing the transition of an elongated BEC (with long range phase
coherence) to a quasi one-dimensional Tonks-Girardeau gas (without phase
coherence), or the transition of a rapidly rotating BEC into the quantum Hall
regime. The XFROG method could also be of interest in the study of the
so-called BCS-BEC crossover where a superfluid of Cooper pairs goes over
into a BEC of tightly bound molecules.

We thank H. Giessen for valuable criticism on the manuscript.
This work has been supported in part by the US Office of Naval
Research, by the National Science Foundation, by the US Army
Research Office, and by the National Aeronautics and Space
Administration.

\bibliography{frog}

\end{document}